\documentclass[12pt,preprint]{aastex}
\usepackage{amssymb,amsmath,epsfig}
\allowdisplaybreaks

\begin{document}

\title{Stability Analysis of Axial Reflection Symmetric Spacetime}

\author{M. Sharif\thanks{msharif.math@pu.edu.pk} and M. Zaeem Ul
Haq Bhatti \thanks{mzaeem.math@pu.edu.pk}}
\affil{Department of Mathematics, University of the Punjab,\\
Quaid-e-Azam Campus, Lahore-54590, Pakistan.}

\begin{abstract}
In this paper, we explore instability regions of non-static axial
reflection symmetric spacetime with anisotropic source in the
interior. We impose linear perturbation on the Einstein field
equations and dynamical equations to establish the collapse
equation. The effects of different physical factors like energy
density and anisotropic stresses on the instability regions are
studied under Newtonian and post-Newtonian limits. We conclude that
stiffness parameter has a significant role in this analysis while
the reflection terms increase instability ranges of non-static axial
collapse.
\end{abstract}

\keywords{Axial symmetry; Relativistic fluids; Stability}

\section{Introduction}

Self-gravitating objects pass through different intense phases of
dynamical activities during the evolution of the star model.
Anisotropy cannot be ignored in the study of rotating stars which is
closely related with axial symmetry. However, there is no exterior
metric which coincides with the sources of such interior. It is
noted that in conventional celestial objects radial and tangential
pressures exist instead of purely isotropic fluids. The theoretical
advances indicate that such objects in which density ranges upto
$\mu<10^{15}gcm^{-3}$ will be anisotropic \citep{1,1a}. \citet{3}
classified spatially homogeneous axially symmetric spacetimes for an
imperfect fluid configuration by keeping constant ratio between
shear and expansion.

\citet{5a} studied an approach to examine the slow adiabatic
contraction of anisotropic spheres. They related the radial and
tangential pressures by a quadratic law in radial coordinate,
defining in this way an anisotropic law (or second equation of
state). \citet{5b} explored solutions of the field equations
depending upon two arbitrary functions, i.e., anisotropic and
generating functions which measure the degree of anisotropy and
relevant physical quantities. The solutions are then matched with
the Schwarzschild exterior metric. \citet{5c} studied the effects of
anisotropic pressure on the properties of gravitationally bound
spherically symmetric object. They found that anisotropic pressure
can have significant effects on the structure and properties of
stellar objects. Particularly, the anisotropy can change the
critical mass and surface redshift of the equilibrium
configurations. \citet{5e} presented an exact analytical treatment
to the field equations describing spherically symmetric anisotropic
matter configuration. They concluded that anisotropic pressure (with
radial pressure obeying linear equation of state) increases the
maximum radius and mass of the quark star, which in their case is
around three solar masses. \citet{5f} have extended the formalism
developed by Chandrasekhar to discuss the significance of
anisotropic pressure on the stability of spherical objects against
radial perturbations in the scenario of Newtonian gravity and
general relativity. They have also discussed this formalism to study
anisotropic spheres with constant energy density and energy
densities of $\frac{1}{r^2}$ profile. Recently, \cite{ab1} explored
spherical collapse and expansion of anisotropic cylindrically
self-gravitating systems with charged background.

\citet{6} explored a class of exact solutions for anisotropic
spherical stars with a physically reasonable form of energy density.
These solutions help to describe anisotropic nature of compact
objects under strong gravitational fields. \citet{7} discussed the
formation of anisotropic compact star with variable cosmological
constant and checked all the regularity conditions as well as
stability of their model. \citet{7a,7b} investigated the role of
different physical factors including anisotropic pressure on
distinct star models.

The stability of self-gravitating stars is an important issue as
only stable equilibrium models are viable. A general relativistic
treatment is required for a precise evaluation of instability
regimes. Stability analysis of self-gravitating stars have been
performed by several authors since the pioneering work of
\citet{10}. The Newtonian (N) and post-Newtonian (pN) approximations
have mainly been used to investigate the structure of a rotating
star in the framework of general relativity \citep{11}. \citet{12}
studied general relativistic treatment for the stability of
axisymmetric spacetime subject to the radial perturbation along
sequence of rotating stars.

\citet{13} investigated gravitational effects for a specific model
and showed that behavior of the system can be drastically changed
due to spacetime curvature leading to stability or instability.
\citet{14} explored the existence and stability of isotropic
homogeneous star subject to perturbations in $f(R)$ gravity.
\citet{15,15a} discussed stability analysis via perturbation without
using equation of state in anisotropic stellar interior whose
results with astrophysical relevance has also been studied
\citep{5f}. Recently, stability analysis for spherical \citep{16a}
as well as cylindrical \citep{16b} configurations are performed by
using radial perturbation in the framework of $f(R)$ gravity in
which they concluded that instability ranges depend only on material
variables with zero expansion independent of the fact that how much
the fluid is stiff. However, it shows dependence on the stiffness
parameter in the presence of expansion scalar.

\citet{18a} showed that instability range is independent of fluid
stiffness with zero expansion which is compatible with the study of
Tolman mass. \citet{19} studied instability of black holes in
massive gravity theory and concluded that linear perturbation around
the simplest black hole leads to unstable mode. \citet{20} found a
connection between gravity and thermodynamics and discussed
stability properties under the influence of cosmological constant.
We have explored some instability regions for self-gravitating
fluids with and without expansion-free condition by imposing linear
perturbation \citep{20a,20b}. \citet{21} used the gauge-invariant
perturbation theory to discuss stability of spherically symmetric
spacetime with anisotropic fluid under axial perturbation. Recently,
\citet{ab2} discussed dynamical properties of commutative black hole
and found its total energy.

\citet{22} systematically derived three classes of axially symmetric
spatially homogeneous spacetime for an ideal fluid using equation of
state. \citet{23} solved the field equations by using anisotropic
features of the universe with axially symmetric spacetime and showed
that their solution represent evolution of the early universe. The
dynamical stability of rotating or axial reflection symmetric stars
against linear perturbations as well as the final fate of collapse
has not been established definitively. A number of interesting
attempts in this direction have already been made with a restricted
class in a modified gravity \citep{24,24a}.

The main idea of this work is to present an analytic treatment in
spirit of Chanrasekhar's work \citep{10} to identify the instability
eras of non-static axial geometry with reflection symmetry filled
with anisotropic matter. The paper is outlined as follows. In the
next section, we present some basic equations for axial spacetime
which are used to develop our analysis. Section \textbf{3} provides
perturbation scheme to all equations obtained in section \textbf{2}
and consequently derives the collapse equation. In section
\textbf{4}, we identify the instability regions with N and pN limits
from the collapse equation. In the last section, we summarize our
results.

\section{Anisotropic Source and Conservation Laws}

We take a non-static axial spacetime with reflection symmetry in
spherical coordinates given by \citep{25,25a}
\begin{equation}\label{1}
ds^2=-A^2(t,r,\theta)dt^{2}+B^2(t,r,\theta)(dr^{2}+r^2d\theta^2)
+2G(t,r,\theta){dt}d\theta+C^2(t,r,\theta)d\phi^2.
\end{equation}
It excludes explicitly the term representing rotation in the
geometry to avoid the complications in the calculations. We consider
anisotropic fluid whose energy-momentum tensor is given as
\begin{equation}\label{2}
T^{(m)}_{\alpha\beta}=(\mu+P)V_\alpha V_\beta+P
g_{\alpha\beta}+\Pi_{\alpha\beta},
\end{equation}
here
\begin{align*}\nonumber
\Pi_{\alpha\beta}&=\frac{1}{3}(\Pi_{II}+2\Pi_I)(K_\alpha
K_\beta-\frac{1}{3}h_{\alpha\beta})+\frac{1}{3}(\Pi_I+2\Pi_{II})(L_\alpha
L_\beta-\frac{1}{3}h_{\alpha\beta})\\\label{7}
&+\Pi_{KL}(K_{\alpha}L_\beta+K_\beta L_\alpha),
\end{align*}
which includes
\begin{align*}\nonumber
&h_{\alpha\beta}=g_{\alpha\beta}+V_\alpha V_\beta,\quad
\Pi_{KL}=K^{\alpha}L^{\beta}T_{\alpha\beta},\quad\Pi_{I}=(2K^\alpha
K^\beta -S^\alpha S^\beta-L^\alpha
L^\beta)T_{\alpha\beta},\\\nonumber
&\Pi_{II}=(2L^{\alpha}L^\beta-K^\alpha K^\beta -S^\alpha
S^\beta)T_{\alpha\beta},
\end{align*}
where $\mu,~\Pi_{\alpha\beta}$ and $h_{\alpha\beta}$ are the energy
density, anisotropic stress tensor and projection tensor,
respectively. Also, $P$ is isotropic pressure and
$\Pi_{I}\neq\Pi_{II}\neq\Pi_{KL}$ indicates anisotropic contribution
of stress tensor. Moreover, $S_\alpha,~L_\alpha,~K_\alpha$ and
$V_\alpha$ are unit four-vectors and four-velocity, respectively
while $\alpha,~\beta$ represent Lorentz indices. In comoving
coordinate system, one can assume these four-vectors as
\begin{equation}\label{3}
S_{\alpha}=C\delta^{3}_{\alpha}\quad
L_{\alpha}=\frac{\sqrt{r^2A^2B^2+G^2}}{A}\delta^{2}_{\alpha},\quad
K_{\alpha}=B\delta^{1}_{\alpha},\quad
V_{\alpha}=-A\delta^{0}_{\alpha}+\frac{G}{A}\delta^{2}_{\alpha},
\end{equation}
satisfying
\begin{align}\nonumber
&K^\alpha L_\alpha=S^\alpha L_\alpha=K^\alpha S_\alpha=V_\alpha
S_\alpha=V^\alpha K_\alpha = V^\alpha L_\alpha=0,\\\nonumber
&K_\alpha K^\alpha=S_\alpha S^\alpha=-V^\alpha V_\alpha=L_\alpha
L^\alpha=1.
\end{align}
Using these unit four-vectors, the non-zero components of
Eq.(\ref{2}) become
\begin{eqnarray}\nonumber
&&T_{00}=\mu A^2,\quad T_{02}=-\mu G,\quad
T_{11}=\left(P+\frac{1}{3}\Pi_I\right)B^2,\\\nonumber
&&T_{22}=\mu\frac{G^2}{A^2}+\left(\frac{r^2A^2B^2+G^2}{A^2}\right)
\left(P+\frac{1}{3}\Pi_{II}\right),\\\label{4}
&&T_{12}=\Pi_{KL}\left(\frac{B}{A}\sqrt{r^2A^2B^2+G^2}\right),\quad
T_{33}=\left[P-\frac{1}{3}\left(\Pi_I+\Pi_{II}\right)\right]C^2.
\end{eqnarray}

In order to describe dynamical nature of any self-gravitating
system, the conservation law, $T^{\alpha\beta}_{~~~;\beta}=0$, has a
crucial role which from Eqs.(\ref{2}) and (\ref{4}) for $\alpha=0,1$
yield the following couple of equations
\begin{align}\nonumber
&\dot{\mu}-\mu\left[\frac{\dot{B}}{B}+\frac{\dot{C}}{C}
+\frac{1}{r^2A^2B^2+G^2}\left\{r^2A\dot{A}B^2+G\dot{G}
+r^2A^2B\dot{B}\right\}\right]+(\mu+P)\\\nonumber
&\times\frac{A^2B^2}
{r^2A^2B^2+G^2}\left[r^2\left(\frac{2\dot{B}}{B}+\frac{\dot{C}}{C}
\right)+\frac{G^2}{A^2B^2}\left(\frac{\dot{B}}{B}
+\frac{\dot{G}}{G}-\frac{\dot{A}}{A}
+\frac{\dot{C}}{C}\right)\right]+\frac{\Pi_I}{3}\\\nonumber
&\times\left(\frac{\dot{B}}{B}-\frac{\dot{C}}{C}\right)
+\frac{\Pi_{II}}{r^2A^2B^2+G^2}\left[r^2A^2B^2\left(\frac{\dot{B}}{B}
-\frac{\dot{C}}{C}\right)+{G^2}\left(\frac{\dot{G}}{G}
-\frac{\dot{A}}{A}-\frac{\dot{C}}{C}\right)\right]\\\label{6}
&=0,\\\nonumber
&P'+\frac{2}{9}\left(2\Pi_I'+\Pi_{II}'\right)+\left[P
+\frac{2}{9}\left(2\Pi_I+\Pi_{II}\right)\right]\left[\frac{C'}{C}
+\frac{3GG'}{2}+\frac{r^2A^2B^2}{r^2A^2B^2+G^2}\right.\\\nonumber
&\times\left.\left(\frac{A'}{A}+\frac{2B'}{B}+\frac{2}{r}
-\frac{(rB)'}{rB}\right)\right]
-\frac{r^2AB^5}{(r^2A^2B^2+G^2)^{\frac{3}{2}}}\Pi_{KL,\theta
}-\frac{r^2AB^5}{(r^2A^2B^2+G^2)^{\frac{3}{2}}}
\\\nonumber &\times\left\{\frac{A_\theta}{A}\right.
\left.+\frac{6B_\theta}{B}+\frac{C_\theta}{C}+\frac{4GG_\theta}
{r^2A^2B^2+G^2}+\frac{4r^2A^2B^2}{r^2A^2B^2+G^2}\left(\frac{A_\theta}{A}
+\frac{B_\theta}{B}\right)\right\}\Pi_{KL}\\\nonumber &+\frac{\mu
r^4A^4B^4}{(r^2A^2B^2+G^2)^2}\left(B\dot{B}+\frac{A'}{A}
-\frac{GA_\theta}{r^2AB^2}\right)
-\frac{\mu~r^2A^2G^2B^2}{(r^2A^2B^2+G^2)^2}\left(\frac{G'}{2G}
+\frac{(rB)'}{rB}\right)\\\label{7} &=0,
\end{align}
where dot and prime stand for differentiation with respect to $t$
and $r$ while subscript $\theta$ indicates $\theta$-differentiation.

\section{Perturbation Approach}

Here we use perturbation technique to obtain perturbed form of all
the previous equations up to first order keeping the perturbation
parameter $\varepsilon$ in the interval $(0,1)$. The initial
configuration of the system is considered to be static while after
perturbation it enters into non-static phase with the same time
dependence of metric coefficients. Consequently, the metric and
matter variables are perturbed as follows \citep{18a,16a,16b}
\begin{eqnarray}\label{9}
A(t,r,\theta)&=&A_0(r,\theta)+\varepsilon
T(t)a(r,\theta),\\\label{10}
B(t,r,\theta)&=&B_0(r,\theta)+\varepsilon
T(t)b(r,\theta),\\\label{11}
C(t,r,\theta)&=&C_0(r,\theta)+\varepsilon T(t){c}(r,\theta),
\\\label{12} G(t,r,\theta)&=&G_0(r,\theta)+\varepsilon
T(t)g(r,\theta), \\\label{13}
\mu(t,r,\theta)&=&\mu_0(r,\theta)+\varepsilon
\bar{\mu}(t,r,\theta),\\\label{14}
P(t,r,\theta)&=&P_{0}(r,\theta)+\varepsilon
\bar{P}(t,r,\theta),\\\label{15}
\Pi_{I}(t,r,\theta)&=&\Pi_{I0}(r,\theta)+\varepsilon
\bar{\Pi_{I}}(t,r,\theta),\\\label{16}
\Pi_{II}(t,r,\theta)&=&\Pi_{II0}(r,\theta)+\varepsilon
\bar{\Pi_{II}}(t,r,\theta), \\\label{17}
\Pi_{KL}(t,r,\theta)&=&\Pi_{KL0}(r,\theta)+\varepsilon
\bar{\Pi_{KL}}(t,r,\theta).
\end{eqnarray}
Using the above equations, the first conservation law (\ref{6}) is
perturbed as
\begin{align}\nonumber
&\dot{\bar{\mu}}=-\left[\mu_0\left\{\frac{b}{B_0}+\frac{c}{C_0}+\frac{1}{Z_0}
\left(r^2aA_0B_0^2+gG_0+r^2bB_0A_0^2\right)\right\}+(\mu_0+P_0)\right.\\\nonumber
&\times\frac{A_0^2B_0^2}{Z_0}\left\{r^2\left(\frac{2b}{B_0}
+\frac{2c}{C_0}\right)+\frac{G_0^2}{A_0^2B_0^2}
\left(\frac{b}{B_0}+\frac{g}{G_0}-\frac{a}{A_0}+\frac{c}{C_0}\right)\right\}
+\frac{\Pi_{I0}}{3}\\\nonumber
&\times\left.\left(\frac{b}{B_0}-\frac{c}{C_0}\right)
+\frac{\Pi_{II0}}{3Z_0}\left\{r^2A_0^2B_0^2\left(\frac{b}{B_0}
-\frac{c}{C_0}\right)+G_0^2\left(\frac{g}{G_0}-\frac{a}{A_0}
-\frac{c}{C_0}\right)\right\}\right]\dot{T}.\\\label{18}
\end{align}
It is interesting to mention here that only the non-static part of
first conservation law exists while the static part vanishes for
this case. Similarly, the static part of second conservation law
leads to
\begin{align}\nonumber
&P'_0+\frac{2}{9}\left(2\Pi_{I0}'+\Pi_{II0}'\right)+\left[P_0
+\frac{2}{9}\left(2\Pi_{I0}+\Pi_{II0}\right)\right]\left[\frac{C'_0}{C_0}
+\frac{3G_0G'_0}{2}+\frac{r^2A^2_0B^2_0}{Z_0}\right.\\\nonumber
&\times\left.\left(\frac{A'_0}{A_0}+\frac{2B'_0}{B_0}
+\frac{2}{r}-\frac{1}{r}-\frac{B_0'}{B_0}\right)\right]
-\frac{r^2A_0B^5_0}{Z_0^{\frac{3}{2}}}\Pi_{KL0,\theta}
-\frac{r^2A_0B^5_0}{Z_0^{\frac{3}{2}}}
\left\{\frac{A_{0\theta}}{A_0}\right.\\\nonumber &
\left.+\frac{6B_{0\theta}}{B_0}+\frac{C_{0\theta}}{C_0}
+\frac{4G_0G_{_0\theta}}{Z_0}+\frac{4r^2A^2_0B^2_0}{Z_0}
\left(\frac{A_{0\theta}}{A_0}
+\frac{B_{0\theta}}{B_0}\right)\right\}\Pi_{KL0}\\\label{19}
&+\frac{\mu_0r^4A^4_0B^4_0}{Z_0^2}\left(\frac{A'_0}{A_0}
-\frac{G_0A_{0\theta}}{r^2A_0B^2_0}\right) -\frac{\mu_0
r^2A^2_0G^2_0B^2_0}{Z_0^2}\left(\frac{G_0'}{2G_0}
+\frac{1}{r}+\frac{B_0'}{B_0}\right)=0.
\end{align}
The non-static part of the second conservation law after
perturbation turns out to be
\begin{align}\nonumber
&\frac{1}{B_0^2}\left\{\bar{P}'+\frac{2}{9}(2\bar{\Pi}_I'
+\bar{\Pi}_{II}')\right\}+\frac{1}{B_0^2}
\left\{\bar{P}+\frac{2}{9}(2\bar{\Pi}_I+\bar{\Pi}_{II})
\right\}\left\{\frac{C_0'}{C_0}
+\frac{3G_0G_0'}{2}\right.\\\nonumber
&\left.+\frac{r^2A_0^2B_0^2}{Z_0}\left(\frac{A_0'}{A_0}
+\frac{B_0'}{B_0}+\frac{2}{r}
-\frac{1}{r}-\frac{B_0'}{B_0}\right)\right\}-\frac{r^2A_0B_0^3}
{Z^{\frac{3}{2}}_0}\bar{\Pi}_{KL,\theta}
-\bar{\Pi}_{KL}\frac{r^2A_0B_0^3}{Z_0^{\frac{3}{2}}}\\\nonumber
&\times\left\{\frac{A_{0\theta}}{A_0}
+\frac{6B_{0\theta}}{B_0}+\frac{C_{0\theta}}{C_0}+\frac{4G_0G_{_0\theta}}
{Z_0}+\frac{4r^2A^2_0B^2_0}{Z_0}\left(\frac{A_{0\theta}}{A_0}
+\frac{B_{0\theta}}{B_0}\right)\right\}+\frac{\bar{\mu}r^4A_0^4}{Z_0^2}\\\nonumber
&\times\left(\frac{A_0'}{A_0}-\frac{G_0A_{0\theta}}{r^2A_0B_0^2}\right)
-\bar{\mu}G_0^2r^2\frac{A_0^2}{Z_0^2}
\left\{\frac{G_0}{2}+\frac{1}{r}+\frac{B_0'}{B_0}\right\}-T\left[
\frac{2b}{B_0^3}\left\{{P_0}'+\frac{2}{9}\right.\right.\\\nonumber
&\times\left.(2{\Pi}_{I0}'+{\Pi}_{II0}')\right\}-
\left[\left\{\left(\frac{c}{C_0}\right)'+\frac{3G_0G_0'}{2}
\left(\frac{g}{G_0}+\frac{g'}{G_0'}\right)\right\}
+\frac{r^2A_0^2B_0^2}{Z_0^2}\right.\\\nonumber
&\times\left(\frac{2a}{A_0}+\frac{2b}{B_0}-\frac{\bar{Z}}{Z_0}\right)
\left(\frac{A_0'}{A_0}+\frac{B_0'}{B_0}+\frac{2}{r}
-\frac{1}{r}-\frac{B_0'}{B_0}\right)+\frac{r^2A_0^2B_0^2}{Z_0}\\\nonumber
&\times\left.\left\{\left(
\frac{a}{A_0}+\frac{2b}{B_0}\right)'-\left(\frac{b}{B_0}\right)'\right\}\right]
\left\{{P_0}+\frac{2}{9}(2{\Pi}_{I0}+{\Pi}_{II0})\right\}
\frac{1}{B_0^2}\\\nonumber
&-\frac{2b}{B_0^2}\left\{{P_0}+\frac{2}{9}(2{\Pi}_{I0}+{\Pi}_{II0})
\right\}\left\{\frac{C_0'}{C_0}
+\frac{3G_0G_0'}{2}+\frac{r^2A_0^2B_0^2}{Z_0}\left(\frac{A_0'}{A_0}
+\frac{B_0'}{B_0}\right.\right.\\\nonumber &\left.\left.+\frac{2}{r}
-\frac{1}{r}-\frac{B_0'}{B_0}\right)\right\}+\frac{r^2A_0B_0^3}
{Z^{\frac{3}{2}}_0}{\Pi}_{KL0,\theta}
\left(\frac{a}{A_0}+\frac{3b}{B_0}-\frac{3\bar{Z}}{Z_0}\right)
+\frac{r^3A_0B_0^3\Pi_{KL0}}{Z_0^{\frac{3}{2}}}\\\nonumber
&\times\left(\frac{a}{A_0}+\frac{3b}{B_0}-\frac{3\bar{Z}}{Z_0}
\right)\left\{\frac{A_{0\theta}}{A_0}
+\frac{6B_{0\theta}}{B_0}+\frac{C_{0\theta}}{C_0}+\frac{4G_0G_{_0\theta}}
{Z_0}+\frac{4r^2A^2_0B^2_0}{Z_0}\left(\frac{A_{0\theta}}{A_0}
\right.\right.\\\nonumber
&\left.\left.+\frac{B_{0\theta}}{B_0}\right)\right\}
+\frac{r^3A_0B_0^3}{Z^{\frac{3}{2}}_0}{\Pi}_{KL0}
\left[\frac{6B_{0\theta}}{B_0}\left(\frac{b_\theta}{B_{0\theta}}
+\frac{b}{B_0}\right)\right.\left(\frac{a}{A_0}+\frac{c}{C_0}\right)_\theta
+\frac{4G_0G_{0\theta}}{Z_0}\\\nonumber
&\times\left.\left(\frac{g}{G_0}+\frac{g_\theta}{G_{0\theta}}
-\frac{\bar{Z}}{Z_0}\right)
+\frac{4r^2A_0^2B_0^2}{Z_0}\left(\frac{2a}{A_0}+\frac{2b}{B_0}
-\frac{\bar{Z}}{Z_0}\right)\left(\frac{a}{A_0}
+\frac{b}{B_0}\right)_\theta\right]\\\nonumber
&-\frac{\mu_0r^4A_0^4}{Z_0^2}\left\{\left(
\frac{a}{A_0}\right)'-\frac{G_0}{r^2}\frac{A_{0\theta}}{A_0B_0^2}
\left(\frac{g}{G_0}+\frac{a_\theta}{A_{0\theta}}-\frac{a}{A_0}
-\frac{2b}{B_0}\right)\right\}\\\label{20}
&\left.+\frac{\mu_0G_0^2A_0^2r^2}{Z_0^2}
\left(\frac{2g}{G_0}+\frac{2a}{A_0}-\frac{2\bar{Z}}{Z_0}\right)
+\frac{\mu_0G_0^2A_0^2r^2}{Z_0^2}
\left\{\frac{g}{2}+\left(\frac{b}{B_0}\right)'\right\}\right]=0.
\end{align}
The non-static part of $02$-component of the field equation as
obtained in Eq.(\ref{21}) can be written as
\begin{equation*}
l\ddot{T}+m\dot{T}+nT=0,
\end{equation*}
here $l,~m$ and $n$ are functions of $r$ and $\theta$. The solutions
of this equation involve stable and unstable configurations. To
describe the instability range, we concentrate over the unstable
part which has the following form \citep{18a,16a,16b,7a,7b}
\begin{equation}\label{22}
T(t)=-\exp(\sqrt{\alpha}t),\quad\textmd{where}\quad
\alpha=\frac{-m+\sqrt{m^2-4ln}}{2l}.
\end{equation}
For the solution to be real, we take $\alpha>0$ with certain
constraints that $m<0$ while $l,~n>0$. Such a solution describes
static system which undergoes collapse with large past time.

Now, we calculate non-static part of anisotropic stresses in terms
of static matter profiles to evaluate the instability regimes. We
take the equation of state introduced by \citet{26} which relates
pressure with energy density using adiabatic index $\Gamma$ as
follows
\begin{equation}\label{23}
\bar{P}=\Gamma\frac{P_{0}}{\mu_0+P_{0}}\bar{\mu},
\end{equation}
where $\Gamma$ represents rigidity or stiffness in the fluid which
is taken to be constant in our stability analysis. The value of
$\bar{\mu}$ from the non-static part of first conservation law after
perturbation can be obtained by integrating Eq.(\ref{18}) with
respect to $t$ as follows
\begin{align*}
&\bar{\mu}=-\left[\mu_0\left\{\frac{b}{B_0}+\frac{c}{C_0}+\frac{1}{Z_0}
\left(r^2aA_0B_0^2+gG_0+r^2bB_0A_0^2\right)\right\}+(\mu_0+P_0)\right.\\\nonumber
&\times\frac{A_0^2B_0^2}{Z_0}
\left\{r^2\left(\frac{2b}{B_0}+\frac{2c}{C_0}\right)+\frac{G_0^2}{A_0^2B_0^2}
\left(\frac{b}{B_0}+\frac{g}{G_0}-\frac{a}{A_0}+\frac{c}{C_0}\right)\right\}
+\frac{\Pi_{I0}}{3}\\\nonumber
&\times\left.\left(\frac{b}{B_0}-\frac{c}{C_0}\right)
+\frac{\Pi_{II0}}{3Z_0}\left\{r^2A_0^2B_0^2\left(\frac{b}{B_0}
-\frac{c}{C_0}\right)+G_0^2\left(\frac{g}{G_0}-\frac{a}{A_0}
-\frac{c}{C_0}\right)\right\}\right]T.
\end{align*}
Using this value of $\bar{\mu}$ in Eq.(\ref{23}), we obtain
\begin{eqnarray}\nonumber
\bar{\Pi}_I&=&-\Gamma\frac{\Pi_{I0}}{\mu_0+\Pi_{I0}}\chi T,\quad
\bar{\Pi}_{II}=-\Gamma\frac{\Pi_{II0}}{\mu_0+\Pi_{II0}}\chi
T,\\\label{24}
\bar{\Pi}_{KL}&=&-\Gamma\frac{\Pi_{KL0}}{\mu_0+\Pi_{KL0}}\chi
T,\quad \bar{P}=-\Gamma\frac{P_{0}}{\mu_0+P_{0}}\chi T,
\end{eqnarray}
here
\begin{align*}
&\chi=\mu_0\left\{\frac{b}{B_0}+\frac{c}{C_0}+\frac{1}{Z_0}
\left(r^2aA_0B_0^2+gG_0+r^2bB_0A_0^2\right)\right\}+(\mu_0+P_0)\\\nonumber
&\times\frac{A_0^2B_0^2}{Z_0}
\left\{r^2\left(\frac{2b}{B_0}+\frac{2c}{C_0}\right)
+\frac{G_0^2}{A_0^2B_0^2}\left(\frac{b}{B_0}+\frac{g}{G_0}-\frac{a}{A_0}+\frac{c}{C_0}\right)\right\}
+\frac{\Pi_{I0}}{3}\\\nonumber
&\times\left(\frac{b}{B_0}-\frac{c}{C_0}\right)
+\frac{\Pi_{II0}}{3Z_0}\left\{r^2A_0^2B_0^2\left(\frac{b}{B_0}
-\frac{c}{C_0}\right)+G_0^2\left(\frac{g}{G_0}-\frac{a}{A_0}
-\frac{c}{C_0}\right)\right\}.
\end{align*}
Substituting these values in Eq.(\ref{20}), it follows that
\begin{align}\nonumber
&-\frac{T\Gamma}{B_0^2}\left(\frac{P_0\chi}{\mu_0+P_0}
+\frac{4\Pi_{I0}\chi}{9(\mu_0+\Pi_{I0})}+
\frac{2\Pi_{II0}\chi}{9(\mu_0+\Pi_{II0})}\right)'
-\left\{\frac{P_0}{\mu_0+P_0}+\frac{4\Pi_{I0}}
{9(\mu_0+\Pi_{I0})}\right.\\\nonumber
&\left.+\frac{2\Pi_{II0}}{9(\mu_0+\Pi_{II0})}\right\}\frac{\chi\Gamma
T}{B_0^2}\left\{\frac{C_0'}{C_0}
+\frac{3G_0G_0'}{2Z_0}\right.\left.+\frac{r^2A_0^2B_0^2}{Z_0}\left(\frac{A_0'}{A_0}
+\frac{1}{r}\right)\right\}+\frac{Tr^2A_0B_0^3\Gamma}{Z^{\frac{3}{2}}_0}\\\nonumber
&\times\left(\frac{\Pi_{KL0}\chi}{\mu_0+\Pi_{KL0}}\right)_\theta
+\Gamma\frac{\Pi_{II0}\chi}{\mu_0+\Pi_{II0}}
\frac{Tr^2A_0B_0^3}{Z_0^{\frac{3}{2}}}\left\{\frac{A_{0\theta}}{A_0}
+\frac{6B_{0\theta}}{B_0}\right.+\frac{C_{0\theta}}{C_0}
+\frac{4G_0G_{_0\theta}}{Z_0}\\\nonumber
&\left.+\frac{4r^2A_0B^2_0}{Z_0}\left(\frac{A_{0\theta}}{A_0}
+\frac{B_{0\theta}}{B_0}\right)\right\}=T\left[\frac{\chi
r^4A_0^4}{Z_0^2}\left(\frac{A_0'}{A_0}-\frac{G_0A_{0\theta}}
{r^2A_0B_0^2}\right)\right.\frac{2b}{B_0^3}\\\nonumber
&\times\left\{{P_0}'+\frac{2}{9}(2{\Pi}_{I0}'+{\Pi}_{II0}')\right\}-
\left[\left\{\left(\frac{c}{C_0}\right)'+\frac{3G_0G_0'}{2Z_0}
\left(\frac{g}{G_0}+\frac{g'}{G_0'} -\frac{\bar{Z}}{Z_0}\right)
\right\}\right.\\\nonumber
&+\frac{r^2A_0^2B_0^2}{Z_0^2}\left(\frac{2a}{A_0}+\frac{2b}{B_0}
-\frac{\bar{Z}}{Z_0}\right)\left(\frac{A_0'}{A_0}+\frac{1}{r}
\right)+\frac{r^2A_0^2B_0^2}{Z_0}\left.\left(\frac{a}{A_0}
+\frac{b}{B_0}\right)'\right]\\\nonumber
&\times\left\{{P_0}+\frac{2}{9}(2{\Pi}_{I0}+{\Pi}_{II0})
\right\}\frac{1}{B_0^2}-\frac{2b}{B_0^2}
\left\{{P_0}+\frac{2}{9}(2{\Pi}_{I0}+{\Pi}_{II0})\right\}
\left\{\frac{C_0'}{C_0}+\frac{3G_0G_0'}{2Z_0}\right.\\\nonumber
&\left.+\frac{r^2A_0^2B_0^2}{Z_0}\left(\frac{A_0'}{A_0}\right.\right.
\left.\left.+\frac{1}{r}
\right)\right\}+\frac{r^2A_0B_0^3}{Z^{\frac{3}{2}}_0}{\Pi}_{KL0,\theta}
\left(\frac{a}{A_0}+\frac{3b}{B_0}-\frac{3\bar{Z}}{Z_0}\right)
+\frac{r^2A_0B_0^3\Pi_{KL0}}{Z_0^{\frac{3}{2}}}\\\nonumber
&\times\left(\frac{a}{A_0}+\frac{3b}{B_0}
-\frac{3\bar{Z}}{Z_0}\right)\left\{\frac{A_{0\theta}}{A_0}
+\frac{6B_{0\theta}}{B_0}+\frac{C_{0\theta}}{C_0}
+\frac{4G_0G_{_0\theta}}{Z_0}+\frac{4r^2A^2_0B^2_0}{Z_0}
\left(\frac{A_{0\theta}}{A_0}\right.\right.\\\nonumber
&\left.\left.+\frac{B_{0\theta}}{B_0}\right)\right\}
+\frac{r^2A_0B_0^3}{Z^{\frac{3}{2}}_0}{\Pi}_{KL0}
\left[\frac{6B_{0\theta}}{B_0}\left(\frac{b_\theta}{B_{0\theta}}
+\frac{b}{B_0}\right)\right.\left(\frac{a}{A_0}
+\frac{c}{C_0}\right)_\theta+\frac{4G_0G_{0\theta}}{Z_0}\\\nonumber
&\times\left.\left(\frac{g}{G_0}+\frac{g_\theta}{G_{0\theta}}
-\frac{\bar{Z}}{Z_0}\right)+\frac{4r^2A_0^2B_0^2}{Z_0}
\left(\frac{2a}{A_0}+\frac{2b}{B_0}-\frac{\bar{Z}}{Z_0}\right)
\left(\frac{a}{A_0}+\frac{b}{B_0}\right)_\theta\right]\\\nonumber
&-\frac{\mu_0r^4A_0^4}{Z_0^2}\left\{\left(
\frac{a}{A_0}\right)'-\frac{G_0}{r^2}\frac{A_{0\theta}}{A_0B_0^2}
\left(\frac{g}{G_0}+\frac{a_\theta}{A_{0\theta}}-\frac{a}{A_0}
-\frac{2b}{B_0}\right)\right\}+\frac{\mu_0G_0^2A_0^2r^2}{Z_0^2}\\\nonumber
&\left.\times\left(\frac{2g}{G_0}+\frac{2a}{A_0}-\frac{2\bar{Z}}{Z_0}\right)
+\frac{\mu_0G_0^2A_0^2r^2}{Z_0^2}\left(\frac{g}{G_0}+\frac{b}{B_0}\right)'\right.
-G_0^2r^2\frac{A_0^2}{Z_0^2}\chi\\\label{25}
&\left.\times\left\{\frac{G_0'}{2G_0}+\frac{1}{r}+\frac{B_0'}{B_0}\right\}\right],
\end{align}
where
\begin{eqnarray}\label{26}
Z_0=r^2A_0^2B_0^2+G_0^2,\quad
\bar{Z}=2r^2A_0^2B_0^2\left(\frac{a}{A_0}+\frac{b}{B_0}\right)+2gG_0.
\end{eqnarray}
This is the required collapse equation with the constraints
$P_0',~\Pi_{I0}',~\Pi_{II0}'<0$ which is very useful to investigate
the instability regions for our systematic analysis.

\section{Vorticity Tensor and Instability Regions}

This section investigates vorticity tensor and dynamical instability
ranges for non-static axial spacetime with the help of equations
obtained in the previous section particularly the collapse equation
subject to N and pN limits. The role of stiffness parameter and its
dependence on physical factors are also analyzed in this scenario.

The kinematical variable responsible for producing local spinning
action of anisotropic fluid configurations is the vorticity tensor.
For reflection axisymmetric spacetime, this tensor in terms of four
vectors, $K_\alpha$ and $L_\alpha$, can be expressed as
\begin{equation}\nonumber
\Omega_{\alpha\beta}=\Omega(K_\beta L_\alpha-L_\beta K_\alpha),
\end{equation}
where
\begin{eqnarray}\nonumber
\Omega=\frac{G}{2B\sqrt{Z}}\left(\frac{G'}{G}-\frac{2A'}{A}\right).
\end{eqnarray}
For the vanishing of vorticity scalar, either $G=0$ or
$\frac{G'}{G}-\frac{2A'}{A}=0$. If we take $G=0$ then it leads to
the vanishing of vorticity scalar. On the other hand, if we take
$\frac{G'}{G}-\frac{2A'}{A}=0$, then it vanishes the metric
coefficient $A(t,r,\theta)$ describing  the temporal component of
the spacetime as follows
\begin{eqnarray}\nonumber
\frac{G'}{G}-\frac{2A'}{A}=0,\\\nonumber
\ln\left(\frac{G\tilde{C}}{A^2}\right)=0,
\end{eqnarray}
where $\tilde{C}=\tilde{C}(t,\theta)$ is an arbitrary function of
integration. Consequently, $G\tilde{C}=A^2$, which implies that for
$G=0$, we have $A=0$ disturbing the existence of our non-static
axial spacetime. Hence,
$$\frac{G'}{G}-\frac{2A'}{A}\neq0.$$
Thus, we take $G=0$ with regularity condition at the center
indicating that vorticity of axisymmetric spacetime exists if and
only if its reflection degrees of freedom exist or more precisely
$\Omega=0\Leftrightarrow G=0$. Consequently the assumption
$\Omega=0$ in the dynamical evolution of non-static axisymmetric
anisotropic metric gives zero value to non-diagonal scale factor,
$G$, whose dynamics has already been discussed in GR \citep{n7b} as
well as in modified gravity theory \citep{24}.

\subsection{Newtonian Limit}

For N limit, we take $A_0=1,~B_0=1,~C_0=r,~G_0=r$ so that $Z_0$
turns out to be $r^2$ for the instability analysis. We also discard
the terms of order $\frac{m_0}{r}$, where $m_0$ is the static
profile of the mass function. The physical requirement of the
collapsing matter, i.e., $P_0,~\Pi_{I0},~\Pi_{II0}<0$, is also
imposed in N approximation. By making use of the above mentioned
constraints, the collapse equation yields
\begin{align*}
&T\Gamma\left[\left(3b+\frac{2c}{r}+\frac{g}{r}\right)\left\{{P_0}+\frac{2}{9}
(2{\Pi}_{I0}+{\Pi}_{II0})\right\}\right]'+\frac{11}{4r}\left\{{P_0}+\frac{2}{9}
(2{\Pi}_{I0}+{\Pi}_{II0})\right\}\\\nonumber
&-\left[\frac{\Pi_{KL0}}{2\sqrt{2}r}
\left(3b+\frac{2c}{r}+\frac{g}{r}\right)\right]_\theta=-T\left[
\left\{{P_0}+\frac{2}{9}
(2{\Pi}_{I0}+{\Pi}_{II0})\right\}\left\{\frac{1}{2}(a+b)'\right.\right.\\\nonumber
&\left.\left.+\left(\frac{c}{r}\right)'+\frac{1}{2r}
\left(2a+11b-\frac{\bar{Z}}{2r^2}\right)\right\}
+\frac{\Pi_{KL0}}{2\sqrt{2}}\left[2(a+b)_\theta\left(2a+2b
-\frac{\bar{Z}}{2r^2}\right)\right]\right.\\\nonumber
&\left.\frac{3}{4r}\left(g'+\frac{g}{r}-\frac{\bar{Z}}{2r^2}\right)\left\{{P_0}+\frac{2}{9}
(2{\Pi}_{I0}+{\Pi}_{II0})\right\}\right]+\frac{\mu_0}{4}\left(
2b'-a' +\frac{7}{2r}+\frac{6c}{r}-\frac{\bar{Z}}{r^2}\right).
\end{align*}
The system will be unstable until it satisfies the following
relation
\begin{eqnarray}\label{27}
\Gamma<\frac{\frac{\mu_0}{4}\left(2b'+\frac{6c}{r}+\frac{7}{2r}-a'\right)+\mathcal{G}_2+\mathcal{A}_1}
{\eta'+\frac{11}{4r}\eta+\mathcal{G}_1-\frac{\Pi_{KL0,\theta}}{2\sqrt{2}r}
\left(3b+\frac{2c}{r}\right)_\theta},
\end{eqnarray}
where
\begin{eqnarray}\nonumber
\eta=\left(3b+\frac{2c}{r}+\frac{g}{r}\right)\xi_1,
\end{eqnarray}
while remaining quantities are defined in Appendix \textbf{A}. It is
well-known that instability will emerge as long as all the terms
given in the above inequality are positive. For this purpose, we
need to take $|\mathcal{A}_1|,~|\mathcal{G}_1|$ and
$|\mathcal{G}_2|$ instead of $\mathcal{A}_1,~\mathcal{G}_1$ and
$\mathcal{G}_2$. We find that the adiabatic index depends on the
static profile of matter variables of axial geometry.

\begin{itemize}
\item In the above instability constraint, the quantities $|\mathcal{G}_1|$ and
$|\mathcal{G}_2|$ incorporate meridional effects that arise due to
non-zero vorticity vector of the collapsing system. It is well-known
from the work of \cite{25} that invoking of reflection effects in
axially symmetric anisotropic stellar object causes the emission of
gravitational radiations. These radiations induce the loss of both
energy and angular momentum, which consequently boosts up the
instability of the reflectional axisymmetric body.
\item  The quantity $\mathcal{A}_1$ includes anisotropic contribution of
axial geometry. It is seen from the expression (\ref{27}) that
anisotropy tends to produce complications in understanding its role
in the stability of axial systems. However, if one considers
positivity of all terms in denominator and numerator of the above
expression then it is seen from (\ref{27}) that anisotropic pressure
tends to increase instability regions. This result is
well-consistent with \citet{15}.
\end{itemize}

\subsection{Post-Newtonian Limit}

For the instability era in the pN approximation, we assume
\begin{eqnarray}\label{28}
A_0=1-\frac{m_0}{r_1},\quad B_0=1+\frac{m_0}{r_1},
\end{eqnarray}
and the terms of the order $\frac{m_0}{r}$ while discarding the
terms containing higher orders of $\frac{m_0}{r}$. The system will
be unstable in the pN region if it satisfies the inequality
\begin{eqnarray}\label{29}
\Gamma<\frac{\frac{\mu_0}{4}\sigma\zeta+\mathcal{G}_3+\mathcal{A}_2+X_1}
{\mathcal{A}_3},
\end{eqnarray}
where static profile terms $X_1$ and $\mathcal{G}_3$ are
non-diagonal and diagonal components of scale factors at pN epoch,
respectively, while $\mathcal{A}_2$ and $\mathcal{A}_3$ incorporate
anisotropic effects in the evolutionary phases of collapsing
self-gravitating axial stellar object. These terms are given in
Appendix \textbf{A}.

\subsubsection{Restricted Class of Anisotropic Axial Spacetime}

On assuming $\mathcal{G}_3=0$, our instability constraint at pN
approximation of axisymmetric object with reflection symmetry
reduces to
\begin{eqnarray}\label{30}
\Gamma<\frac{\frac{\mu_0}{4}\sigma\zeta+\mathcal{A}_2+X_1}
{\mathcal{A}_3}.
\end{eqnarray}
This describes instability range of the restricted class of
non-static axial geometry since it excludes explicitly rotations
around the symmetry axis, i.e., $dtd\phi$ as well as the reflection
terms. This result coincides and supports already calculated
solution \citep{n7b}.

\subsubsection{Reflection and Restricted Class of Isotropic Axial
Spacetime}

On taking equal all principal stresses as well as zero value to
$\mathcal{G}_3$, one can find instability regions of restricted
class of isotropic axisymmetric spacetime from expression
(\ref{29}). However, apart from that by assuming only first of above
limits, one can get dynamical instability constraint of reflection
axisymmetric compatible with perfect fluid. All possible stellar
models of reflection axial symmetric system coupled with perfect
(isotropic) matter configurations have have been explored in detail
by \citet{n123}.

We see that the adiabatic index $\Gamma$ plays a central role to
investigate dynamical instability of the relativistic system. It is
worth mentioning that for $\Gamma<\frac{4}{3}$ and $\Gamma<1$, the
spherical and cylindrical relativistic objects become unstable
respectively thereby enforces the importance of index $\Gamma$.
Infact, the adiabatic index also known as stiffness parameter
demonstrates how much relativistic fluid is stiff. We have
established the relevance of such index in the dynamical instability
as seen from expressions (\ref{27}) and (\ref{29}) depending upon
the static profile of the structural properties of the system. The
system would be in complete hydrostatic equilibrium, if (during
evolution) adiabatic index is able to attain value equal to the
right hand side of the expressions given in (\ref{27}) and
(\ref{29}). However, if stiffness parameter attains a value greater
than the right hand side of expressions (\ref{27}) and (\ref{29}),
then the relativistic system begins to move in the stable window,
thereby ceasing the collapsing mechanism.

\section{Conclusions}

This paper is devoted to investigate dynamical instability of
non-static axially symmetric spacetime by choosing reflection term
in the geometry. Since rotating stars are more stable than
non-rotating, so for the instability regions, we have neglected the
term representing rotation in the general non-static axial
spacetime. It is worth mentioning that for axially symmetric sources
perfect fluid distribution seems to be inflexible restriction, even
in the static case. On the other hand, Bondi coordinates are known
to be very useful for the treatment of gravitational radiation in
vacuum, but are not particularly suitable within the source. An
analytical approach, which shares some similarities with ours,
although restricted to the perfect fluid case, can be found in the
literature. Therefore, here, we have considered a source which
includes all non–vanishing stresses compatible with the symmetry of
the problem to carry out our systematic analysis.

We have explored the field equations and corresponding conservation
laws in this scenario. We have found three independent components
from the conservation law while there exist only two components in
the case of spherical and cylindrical spacetimes
\citep{18a,7a,7b,20a,20b}. The radial perturbation is used for
metric as well as material variables to obtain perturbed form of
these dynamical equations. We have explored static and non-static
parts of independent components of the conservation law. It is found
that only non-static part for the first conservation law exists and
static part vanishes while the remaining equations have both static
as well as non-static components. Using $02$-component of the field
equations with perturbation technique, we have found a solution
which corresponds to both stable and unstable configurations and
start collapsing at large past time diminishing its areal radius
\citep{16a,16b,24, 24a}.

We have developed a general collapse equation to examine the
instability regions using non-static parts of anisotropic stresses
and the solution (\ref{22}). We have explored two instability ranges
under N as well as pN limits and found that instability range is
defined by the adiabatic index \citep{24,24a} unlike expansion-free
case [where it has no role \citep{18a}]. The adiabatic index depends
upon static profile of the energy density, anisotropic pressure and
the reflection term in the spacetime. We conclude that reflection
symmetry increases the unstable range of the axial geometry. The
system will remain unstable until it satisfies the relations
(\ref{27}) and (\ref{29}) while their violation will lead to stable
configuration of the model. It would be interesting to examine the
role of dissipative terms like heat flux on the stability of
non-static axial geometry.

\vspace{0.3cm}

\renewcommand{\theequation}{A\arabic{equation}}
\setcounter{equation}{0}
\section*{Appendix A}

The $02$-component of the Einstein tensor corresponding to our line
element in Eq.(\ref{1}) takes the form
\begin{eqnarray}\nonumber
&&G_{02}=-\frac{1}{4(r^2A^2B^2+G^2)^2}\left[4G^4\left\{\frac{\dot{C}_\theta}{C}
+\frac{\dot{B}_\theta}{B}+\frac{\dot{B}C_\theta}{BC}
+\frac{\dot{C}B_\theta}{CB}\right\}\right.\\\nonumber
&&+4r^4A^4B^2G\left\{\frac{A'C'}{AC}-\frac{B'^2}{B^2}-\frac{2A'B'}{AB}
+\frac{1}{r}\left(\frac{C'}{C}+\frac{G'}{G}-\frac{A'}{A}\right)
-\frac{G''}{2G}+\frac{A'G'}{2AG}\right.\\\nonumber
&&\left.+\frac{G'B'}{GB}-\frac{G'C'}{GC}+\frac{A''}{A}+\frac{B''}{B}
+\frac{C''}{C}+\frac{1}{r^2}\left(\frac{B_{\theta\theta}}{B}
+\frac{C_\theta}{C}\right)\right\}+4r^2A^2G^3\\\nonumber
&&\times\left\{\frac{A'C'}{AC}-\frac{3G'B'}{2GB}+\frac{3G'^2}{4G^2}
-\frac{B'C'}{BC}+\frac{A'^2}{A^2}-\frac{3A'G'}{2AG}+\frac{A''}{A}
+\frac{B''}{B}+\frac{2C''}{C}\right.\\\nonumber
&&+\frac{1}{r}\left(\frac{3B'}{B}+\frac{A_\theta
C_\theta}{AC}+\frac{A'}{A}+\frac{3B'}{B}+\frac{C'}{C}-\frac{3G'}{2G}\right)
+\frac{1}{r^2}\left(\frac{A_\theta B_\theta}{AB}-\frac{B_\theta
G_\theta}{BG}\right.\\\nonumber &&\left.\left.-\frac{B_\theta
C_\theta}{BC}-\frac{G_\theta
C_\theta}{GC}+\frac{C_{\theta\theta}}{C}
+\frac{B_{\theta\theta}}{B}\right)\right\}
+4r^4A^2B^4G\left\{\frac{\dot{A}\dot{B}}{AB}+\frac{\dot{A}\dot{C}}{AC}
-\frac{\dot{B}\dot{C}}{BC}\right.\\\nonumber
&&\left.-\frac{C\ddot{}}{C}-\frac{\ddot{B}}{B}\right\}
+4r^2A^2B^2G^2\left\{\frac{B_\theta
\dot{G}}{BG}-\frac{\dot{A}B_\theta}{AB}-\frac{\dot{B}B_\theta}{B^2}
+\frac{C_\theta\dot{G}}{CG}+\frac{\dot{C}G_\theta}{CG}
-\frac{\dot{A}C_\theta}{AC}\right.\\\nonumber
&&\left.+\frac{\dot{B}G_\theta}{BG}+\frac{\dot{B}C_\theta}
{BC}\right\}+4r^4A^4B^4
\left\{\frac{\dot{B}A_\theta}{BA}-\frac{\dot{B}_\theta}{B}
+\frac{\dot{B}B_\theta}{B^2}
+\frac{A_\theta\dot{C}}{AC}+\frac{\dot{B}C_\theta}{BC}
-\frac{\dot{C}_\theta}{C}\right\}\\\nonumber
&&+4r^2B^2G^3\left\{\frac{\dot{G}\dot{C}}{GC}-\frac{\dot{B}^2}{B^2}
+\frac{4\dot{B}\dot{G}}{BG}-\frac{2\dot{B}\dot{C}}{BC}-\frac{\ddot{C}}{C}
-\frac{\ddot{B}}{B}\right\}+\frac{4G^5}{B^2}\left\{\frac{G''}{G}
-\frac{G'^2}{G^2}\right.\\\label{5}
&&\left.\left.-\frac{2B'C'}{BC}+\frac{G'C'}{2GC}+\frac{C''}{C}
-\frac{G'B'}{GB}\right\}\right].
\end{eqnarray}
For $\alpha=3$, the conservation law,
$T^{\alpha\beta}_{~~~;\beta}=0$, leads to the following equation
\begin{align}\nonumber
&\frac{\mu r^2A^2B^2G}{(r^2A^2B^2+G^2)^2}\left[\frac{\dot{\mu}}{\mu}
+\frac{\dot{A}}{A}+\frac{3\dot{B}}{B}+\frac{\dot{G}}{G}
+\frac{\dot{C}}{C}+\frac{1}{r^2B^2}\left(\frac{\mu_{\theta}}{\mu}
+\frac{2G_\theta}{G}+\frac{2A_\theta}{A}\right)\right.\\\nonumber
&+\frac{1}{r^2A^2B^2+G^2}\left\{4r^2A^2\left(\frac{\dot{A}}{A}
+\frac{\dot{B}}{B}\right)-\frac{4\dot{G}}{G}-GA^2
\left(\frac{5A_\theta}{A}+\frac{2B_\theta}{B}\right)\right.\\\nonumber
&\left.\left.+r^2A^2B^2\left(\frac{\dot{G}}{G}+\frac{\dot{B}}{B}\right)
+\frac{r^2A^3B^2A_\theta}{G}\right\}
-\frac{4G^2G_\theta(r^2A^2B^2+G^2)}{r^2B^2}\right]\\\nonumber
&+\frac{\mu~A^2G^2}{(r^2A^2B^2+G^2)^2}\left\{\frac{B_\theta}{B}
+\frac{C_\theta}{C}-\frac{r^2BG\dot{B}}
{r^2A^2B^2+G^2}\right\}-\frac{r^2AB^3\Pi_{KL}}{(r^2A^2B^2+G^2)^{\frac{3}{2}}}\\\nonumber
&\times\left[\frac{\Pi'_{KL}}{\Pi_{KL}}+\frac{3}{r}
+\frac{4B'}{B}+\frac{A'}{A}+\frac{C'}{C}
+\frac{3}{r^2A^2B^2+G^2}\left\{GG'+r^2A^2B^2\right.\right.\\\nonumber
&\times\left.\left. \left(
\frac{3}{r}+\frac{2A'}{A}+\frac{3B'}{B}\right)
\right\}+\frac{7GG'}{r^2A^2B^2+G^2}\right]+\left\{P+\frac{2}{9}
(\Pi_I+2\Pi_{II})\right\}\\\nonumber
&\times\frac{1}{r^2A^2B^2+G^2}\left[\frac{r^2A^2B^2}
{r^2A^2B^2+G^2}\left\{(2A^2+A)\left(
\frac{A_\theta}{A}\right.\right.\right.\left.+\frac{B_\theta}{B}\right)
-G\left(\frac{\dot{B}}{B}\right)\\\nonumber
&\left.+\frac{2AB_\theta}{B}\right\}+2AA_\theta
+\frac{A^2C_\theta}{C}-\frac{r^2BG\dot{B}}{r^2A^2B^2+G^2}
\left.-\frac{2A^2GG_\theta}{r^2A^2B^2+G^2}-\frac{G\dot{B}}{B}\right]\\\nonumber
&-\frac{P}{C(r^2A^2B^2+G^2)}(G\dot{C}
+A^2C_\theta)+\frac{A^2}{r^2A^2B^2+G^2}\\\label{8}
&\times\left\{P_\theta+\frac{2}{9}(\Pi_{I,\theta}+2\Pi_{II,\theta})\right\}=0.
\end{align}
Using Eqs.(\ref{4}), (\ref{5}) and (\ref{9})-(\ref{13}), the
non-static part of the $02$-component of the field equations,
$G_{\alpha\beta}=8\pi T_{\alpha\beta}$, becomes
\begin{align}\nonumber
&-\frac{2r^4G_0A_0^3B_0A_0'B_0'}{Z_0^4}\left(\frac{g}{G_0}
+\frac{3a}{A_0}+\frac{b}{B_0}
+\frac{a'}{A_0}-\frac{2\bar{Z}}{Z_0}\right)T-\frac{3}{2}
\frac{r^2A_0^2G_0^2G_0'B_0'}{B_0Z_0^2}\\\nonumber
&\times\left(\frac{g'}{G_0'}+\frac{b'}{B_0'}+\frac{2a}{A_0}
+\frac{2g}{G_0}-\frac{2\bar{Z}}{Z_0}
-\frac{b}{B_0}\right)T+\frac{G_0B_0A_0^2r^2B_{0\theta}g\dot{T}}{Z_0^2}\\\nonumber
&-\frac{r^2aG_0^2A_0^2B_0^2\dot{T}}
{Z_0^2}-\frac{br^2A_0^2G_0^2B_{0\theta}}{Z_0^2}-\frac{3r^2A_0A_0'G_0'G_0^2}
{Z_0^2}\left(\frac{a}{A_0}
+\frac{a'}{A_0'}+\frac{g'}{G_0'}\right.\\\nonumber
&\left.+\frac{2g}{G_0}-\frac{2\bar{Z}}{Z_0}\right)T +\frac{C_\theta
G_0^4}{C_0Z_0^2}\dot{T}
+\frac{3r^2G_0G_0'^2A_0^2}{Z_0^2}\left(\frac{g}{G_0}+\frac{2g'}{G_0'}
+\frac{2a}{A_0}-\frac{2\bar{Z}}{Z_0}\right)T\\\nonumber
&+\frac{3rA_0^2B_0'G_0^3}{B_0Z_0^2}\left(\frac{2a}{A_0}
+\frac{b'}{B_0'}-\frac{b}{B_0}+\frac{3g\bar{Z}G_0Z_0}{}\right)T
-\frac{A_0^2G_0^2B_{0\theta}G_{0\theta}T}{B_0Z_0^2}\left(
\frac{2a}{A_0}\right.\\\nonumber &\left.+\frac{2g}{G_0}
+\frac{b_\theta}{B_{0\theta}}+\frac{g_\theta}{G_{0\theta}}
-\frac{b^2\bar{Z}}{B_0Z_0}\right)
-\frac{r^2B_0'C_0'A_0^2G_0^3}{B_0C_0Z_0^2}\left(\frac{b'}{B_0'}
+\frac{c'}{C_0'}+\frac{2a}{A_0}+\frac{3g}{G_0}\right.\\\nonumber
&\left.-\frac{b}{B_0}-\frac{c}{C_0}-\frac{2\bar{Z}}{Z_0}\right)T
-\frac{G_0A_0^4r^4B_0'^2}{Z_0^4}\left(\frac{g}{G_0}+\frac{4a}{A_0}
+\frac{2b}{B_0}-\frac{2\bar{Z}}{Z_0}\right)T\\\nonumber
&+\frac{r^4G_0B_0^2A_0^3A_0'C_0'}{C_0Z_0^2}\left(\frac{g}{G_0}
+\frac{2b}{B_0}+\frac{3a}{A_0}+\frac{a'}{A_0'}
+\frac{c'}{C_0'}-\frac{c}{C_0}-\frac{2\bar{Z}}{Z_0}\right)T\\\nonumber
&+\frac{4G_0B_0^4r^2C_{0\theta}g\dot{T}}{C_0Z_0^2}
+\frac{G_0A_0^2B_0^2r^2G_{0\theta}c\dot{T}}{C_0A_0^2}
-\frac{r^2B_0^2aA_0C_{0\theta}\dot{T}}{C_0Z_0^2}
+\frac{r^2A_0A_0'C_0'G_0^3}{C_0Z_0^2}\\\nonumber
&\times\left(\frac{3g}{G_0}+\frac{a'}{A_0'}+\frac{a}{A_0}
+\frac{c'}{C_0'}-\frac{c}{C_0}-\frac{2\bar{Z}}{Z_0}\right)T
+\frac{G_0A_0^2B_0r^2G_{0\theta}b\dot{T}}{Z_0^2}
-\frac{B_0^3b_\theta A_0^2r^4\dot{T}}{Z_0^2}\\\nonumber
&-\frac{G_0^3B_0r^2b\ddot{T}}{Z_0^2}+\frac{r^2A_0'^2G_0^3}{Z_0^2}\left(
\frac{2a'}{A_0'}+\frac{3g}{G_0}-\frac{2\bar{Z}}{Z_0}\right)T
+\frac{A_0^2B_{0\theta}C_{0\theta}G_0^3T}{B_0C_0Z_0^2}
\left(\frac{2a}{A_0}\right.\\\nonumber
&\left.+\frac{b_\theta}{B_{0\theta}}+\frac{c_\theta}{C_{0\theta}}
-\frac{b}{B_0}-\frac{c}{C_0}+\frac{3g}{G_0}-\frac{2\bar{Z}}{Z_0}\right)
-\frac{4A_0^2G_0^2G_{0\theta}C_{0\theta}T}{C_0Z_0^2}
\left(\frac{2a}{A_0}+\frac{2g}{G_0}\right.\\\nonumber
&\left.+\frac{g_\theta}{G_{0\theta}}+\frac{c_\theta}{C_{0\theta}}
-\frac{c}{C_0}-\frac{2\bar{Z}}{Z_0}\right)+\frac{rC_0'G_0^3}{C_0Z_0^2}
\left(\frac{2a}{A_0}+\frac{c'}{C_0'}-\frac{c}{C_0}
+\frac{3g}{G_0}-\frac{2\bar{Z}}{Z_0}\right)T\\\nonumber
&+\frac{G_0A_0^4B_0^2C_0'r^3}{C_0Z_0^2}\left(
\frac{g}{G_0}+\frac{4a}{A_0}+\frac{2b}{B_0}+\frac{c'}{C_0'}
-\frac{c}{C_0}-\frac{2\bar{Z}}{Z_0}\right)T
-\frac{B_0'C_0'G_0^5}{B_0^3C_0Z_0^2}\\\nonumber
&\times\left(\frac{b'}{B_0'}+\frac{c'}{C_0'}+\frac{5g}{G_0}-\frac{3b}{B_0}
-\frac{c}{C_0}-\frac{2\bar{Z}}{Z_0}\right)T+\frac{TA_0^2G_0^3}{Z_0^2}
\left(\frac{2a}{A_0}+\frac{3g}{G_0}
-\frac{2\bar{Z}}{Z_0}\right)\\\nonumber
&+\frac{G_0^5C_{0\theta}b\dot{T}}{B_0C_0Z_0^2}
+\frac{G_0^5B_{0\theta}c\dot{T}}{B_0C_0Z_0^2}
-\frac{G_0^3G_0'}{B_0^2Z_0^2}\left(\frac{3g}{G_0}+\frac{2g'}{G_0'}
-\frac{2\bar{Z}}{Z_0}-\frac{2b}{B_0}\right)T +\frac{b_\theta
G_0^4\dot{T}}{B_0Z_0^2}\\\nonumber
&-\frac{r^4A_0^4B_0^5c_\theta\dot{T}}{Z_0^2}
-\frac{br^2B_0A_0^2G_0^2C_{0\theta}T}{C_0Z_0^2}
+\frac{br^4A_0^2B_0^2B_{0\theta}\dot{T}}{Z_0^2}
+\frac{br^4A_0^3B_0^3A_{0\theta}\dot{T}}{Z_0^2}\\\nonumber
&+\frac{r^4A_0^3B_0^2A_0'G_0'T}{2G_0^2}\left(\frac{2b}{B_0}
+\frac{3a}{A_0}+\frac{a'}{A_0'}+\frac{g}{G_0'}
-\frac{2\bar{Z}}{Z_0}\right)+\frac{4r^4B_0A_0^4B_0'G_0'T}{Z_0^2}\\\nonumber
&\times\left(\frac{3g}{G_0}+\frac{a}{A_0}
+\frac{a_\theta}{A_{0\theta}}+\frac{b_\theta}{B_{0\theta}}
-\frac{b}{B_0}\right)-\frac{3rA_0^2G_0^2G_0'}{2Z_0^2}
\left(\frac{2a}{A_0}+\frac{g'}{G_0'}+\frac{2g}{G_0}
-\frac{2\bar{Z}}{Z_0}\right)\\\nonumber
&+\frac{2r^3A_0^4B_0^2G_0'T}{Z_0^2}
\left(\frac{4a}{A_0}+\frac{2b}{B_0}+\frac{g'}{G_0'}
-\frac{2\bar{Z}}{Z_0}\right)
+\frac{cr^4B_0^4A_0^3A_{0\theta}\dot{T}}{C_0Z_0^2}
+\frac{G_0^3A_0A_{0\theta}C_{0\theta}}{C_0Z_0^2}\\\nonumber
&\times\left(\frac{3g}{G_0}+\frac{a}{A_0}+\frac{a_\theta}{A_{0\theta}}
+\frac{c_\theta}{C_{0\theta}}-\frac{c}{C_0}-\frac{2\bar{Z}}{Z_0}\right)T
-\frac{r^4A_0^4B_0^2G_0'C_0'T}{C_0Z_0^2}\left(
\frac{g'}{G_0'}+\frac{c'}{C_0'}\right.\\\nonumber
&\left.+\frac{4a}{A_0}+\frac{2b}{B_0}-\frac{c}{C_0}
-\frac{2\bar{Z}}{Z_0}\right)+\frac{\dot{T}r^4bA_0^4B_0^3C_{0\theta}}
{C_0Z_0^2}-\frac{TG_0'B_0'G_0^4}{2B_0^3Z_0^2}
\left(\frac{g'}{G_0'}+\frac{b'}{B_0'}+\frac{4g}{G_0}\right.\\\nonumber
&\left.-\frac{3b}{B_0}-\frac{2\bar{Z}}{Z_0}\right)
+\frac{2G_0'C_0'G_0^4}{B_0^2C_0Z_0^2}\left(\frac{g'}{G_0'}
+\frac{c'}{C_0'}+\frac{4g}{G_0}-\frac{2b}{B_0}
-\frac{c}{C_0}-\frac{2\bar{Z}}{Z_0}\right)T\\\nonumber
&-\frac{\ddot{T}r^2cB_0^2G_0^3}{C_0Z_0^2}-\frac{\ddot{T}r^4cA_0^2B_0^4G_0}
{C_0Z_0^2}-\frac{r^4A_0^4B_0^2G_0''T}{2Z_0^2}
\left(\frac{c}{C_0}+\frac{g''}{G_0''}+\frac{2b}{B_0}
+\frac{4a}{A_0}-\frac{2\bar{Z}}{Z_0}\right)\\\nonumber
&+\frac{TG_0''G_0^4}{2Z_0^2B_0^2}\left(\frac{g''}{G_0''}
+\frac{4g}{G_0}-\frac{2\bar{Z}}{Z_0}-\frac{2b}{B_0}\right)
+\frac{TC_0''G_0^5}{Z_0^2B_0^2C_0}
\left(\frac{c''}{C_0''}+\frac{5g}{G_0}-\frac{2b}{B_0}
-\frac{c}{C_0}-\frac{2\bar{Z}}{Z_0}\right)\\\nonumber
&+\frac{A_0^2G_0^3C_{0\theta\theta}}{C_0Z_0^2}\left(\frac{2a}{A_0}
+\frac{c_{\theta\theta}}{C_{0\theta\theta}}
+\frac{3g}{G_0}-\frac{2\bar{Z}}{Z_0}-\frac{c}{C_0}\right)T
+\frac{r^2A_0A_0''G_0^3T}{Z_0^2}
\left(\frac{a}{A_0}\right.\\\nonumber
&\left.+\frac{a''}{A_0''}+\frac{3g}{G_0}-\frac{\bar{Z}}{Z_0}\right)
+\frac{r^2A_0^2B_0''G_0^3T}{B_0Z_0^2}
\left(\frac{2a}{A_0}+\frac{b''}{B_0''}+\frac{3g}{G_0}-\frac{b}{B_0}
-\frac{\bar{Z}}{Z_0}\right)\\\nonumber
&+\frac{r^4A_0^3B_0^2G_0A_0''}{Z_0^2}\left(\frac{g}{G_0}
+\frac{2b}{B_0}+\frac{3a}{A_0}+\frac{a''}{A_0''}
-\frac{\bar{Z}}{Z_0}\right)T+\frac{r^2A_0^4B_0B_{0\theta}}{Z_0^2}
\left(\frac{g}{G_0}\right.\\\nonumber
&\left.+\frac{4a}{A_0}+\frac{b}{B_0}+\frac{b_\theta}{B_{0\theta}}
-\frac{\bar{Z}}{Z_0}\right)T+\frac{G_0A_0^4B_0r^4B_0''}{Z_0^2}
\left(\frac{g}{G_0}+\frac{4a}{A_0}-\frac{b}{B_0}+\frac{b''}{B_0''}
-\frac{\bar{Z}}{Z_0}\right)T\\\nonumber
&+\frac{A_0^2G_0^3B_{0\theta\theta}}{B_0Z_0^2}\left(\frac{2a}{A_0}
+\frac{b_{\theta\theta}}{B_{0\theta\theta}}
+\frac{3g}{G_0}-\frac{b}{B_0}-\frac{\bar{Z}}{Z_0}\right)T
+\frac{2C_0''A_0^2r^2G_0^3}{C_0Z_0^3}\left(
\frac{c''}{C_0''}\right.\\\nonumber&\left.-\frac{\bar{Z}}{Z_0}
-\frac{c}{C_0}+\frac{2a}{A_0}+\frac{3g}{G_0}\right)T
+\frac{G_0C_0''B_0^2A_0^4r^4}{C_0Z_0^2}\left(\frac{g}{G_0}
+\frac{c''}{C_0''}+\frac{2b}{B_0}
+\frac{4a}{A_0}\right.\\\nonumber&\left.-\frac{c}{C_0}
-\frac{2\bar{Z}}{Z_0}\right)T+\frac{4G_0A_0^4B_0^2r^2C_{0\theta\theta}}{C_0Z_0^2}
\left(\frac{g}{G_0}+\frac{4a}{A_0}+\frac{2b}{B_0}
+\frac{c_{\theta\theta}}{C_{0\theta\theta}}-\frac{c}{C_0}
-\frac{\bar{Z}}{Z_0}\right)T\\\nonumber
&=-\left[\mu_0\left\{\frac{b}{B_0}+\frac{c}{C_0}+\frac{1}{Z_0}
\left(r^2aA_0^2B_0^2+gG_0+r^2bB_0G_0\right)\right\}
+(\mu_0+P_0)\right.\\\nonumber
&\times\frac{A_0^2B_0^2}{Z_0^2}\left\{r^2\left(\frac{2b}{B_0}
+\frac{2c}{C_0}\right)+\frac{A_0^2}{A_0^2B_0^2}
\left(\frac{b}{B_0}+\frac{g}{G_0}-\frac{a}{A_0}
+\frac{c}{C_0}\right)\right\}+\frac{\Pi_{I0}}{3}\\\label{21}
&\times\left.\left(\frac{b}{B_0}-\frac{c}{C_0}\right)
+\frac{\Pi_{II0}}{3Z_0}\left\{r^2A_0^2B_0^2\left(\frac{b}{B_0}
-\frac{c}{C_0}\right)+G_0^2\left(\frac{g}{G_0}-\frac{a}{A_0}
-\frac{c}{C_0}\right)\right\}\right]T.
\end{align}
Using pN approximation, Eq.(\ref{25}) takes the form
\begin{align}\nonumber
&-{T\Gamma}\left(1-\frac{2m_0}{r_1}\right)\left(\frac{P_0\chi}{\mu_0+P_0}
+\frac{4\Pi_{I0}\chi}{9(\mu_0+\Pi_{I0})}+
\frac{2\Pi_{II0}\chi}{9(\mu_0+\Pi_{II0})}\right)'
-\left\{\frac{P_0}{\mu_0+P_0}\right.\\\nonumber&
\left.+\frac{4\Pi_{I0}}{9(\mu_0+\Pi_{I0})}+\frac{2\Pi_{II0}}{9(\mu_0
+\Pi_{II0})}\right\}\left(1-\frac{2m_0}{r_1}\right){\chi\Gamma
T}\left\{\frac{7}{4r}+\frac{1}{2}\left(1-\frac{4m_0^2}{r_1^2}\right)\right.\\\nonumber
&\times\left. \left(1-\frac{m_0^2}{r_1^2}
+\frac{1}{r}\right)\right\}+\frac{T\Gamma}{2\sqrt{2}r}
\left(1-\frac{m_0}{r_1}\right)\left(1+\frac{3m_0}{r_1}\right)
\left(\frac{\Pi_{KL0}\chi}{\mu_0+\Pi_{KL0}}\right)_\theta\\\nonumber
&=T\left[\frac{b\chi}{2}\left(1-\frac{4m_0}{r_1}\right)\left(1
+\frac{m_0}{r_1}\right)\left(1-\frac{3m_0}{r_1}\right)\left(1
-\frac{m_0}{r_1}\right)'\right.\\\nonumber
&\times\left\{{P_0}'+\frac{2}{9}(2{\Pi}_{I0}'+{\Pi}_{II0}')\right\}-
\left[\left\{\left(\frac{c}{r}\right)'+\frac{3}{4r}
\left(\frac{g}{r}+{g'}-\frac{\bar{Z}}{2r^2}\right)
\right\}\right.\\\nonumber
&+\frac{1}{2r^2}\left(1-\frac{4m_0^2}{r_1^2}\right)\left(
{2a}\left(1+\frac{m_0}{r_1}\right)+{2b}
\left(1-\frac{m_0}{r_1}\right)
-\frac{\bar{Z}}{2r^2}\right)\left(\left(1-\frac{m_0}{r_1}\right)'\right.\\\nonumber
&\times\left.\left(1+\frac{m_0}{r_1}\right)+\frac{1}{r}
\right)+\frac{1}{2}\left.\left({a}\left(1+\frac{m_0}{r_1}\right)
+{b}\left(1-\frac{m_0}{r_1}\right)\right)'\right]\left\{{P_0}+\frac{2}{9}
(2{\Pi}_{I0}+{\Pi}_{II0})\right\}\\\nonumber
&\times\left(1-\frac{2m_0}{r_1}\right)-{2b}\left(1-\frac{2m_0}{r_1}\right)
\left\{{P_0}+\frac{2}{9}(2{\Pi}_{I0}+{\Pi}_{II0})\right\}
\left\{\frac{7}{4r}\right.+\frac{1}{2}\left(1-\frac{4m_0^2}{r_1^2}\right)\\\nonumber
&\left.\times\left(\left(1-\frac{m_0}{r_1}\right)'\left(1
+\frac{m_0}{r_1}\right)\right.\right.
\left.\left.+\frac{1}{r}\right)\right\}+\frac{1}{2\sqrt{2}r}
{\Pi}_{KL0,\theta}\left(1-\frac{m_0}{r_1}\right)
\left(1+\frac{3m_0}{r_1}\right)\\\nonumber
&\times\left({a}\left(1+\frac{m_0}{r_1}\right)+{3b}\left(1
-\frac{m_0}{r_1}\right)-\frac{3\bar{Z}}{2r^2}\right)
+\frac{\Pi_{KL0}}{2\sqrt{2}r}\left[6\left(1-\frac{m_0}{r_1}
\right)\right.\\\nonumber
&\times\left.\left(1+\frac{m_0}{r_1}\right)_\theta
\left({b_\theta}\left(1-\frac{m_0}{r_1}\right)_\theta
+{b}\left(1-\frac{m_0}{r_1}\right)\right)\right.\left({a}
\left(1+\frac{m_0}{r_1}\right) +\frac{c}{r}\right)_\theta\\\nonumber
&\left.+2\left(1-\frac{4m_0^2}{r_1^2}\right)
\left({2a}\left(1+\frac{m_0}{r_1}\right)+{2b}\left(1
-\frac{m_0}{r_1}\right)-\frac{\bar{Z}}{r^2}\right)
\left({a}\left(1+\frac{m_0}{r_1}\right)\right.\right.\\\nonumber
&+\left.\left.{b}\left(1-\frac{m_0}{r_1}\right)\right)_\theta\right]
-\frac{\mu_0}{4}\left(1-\frac{4m_0}{r_1}\right)\left\{\left(
{a}\left(1+\frac{m_0}{r_1}\right)\right)'-\frac{1}{r}\left(1
-\frac{m_0}{r_1}\right)_\theta\right.\\\nonumber
&\left.\times\left(1+\frac{m_0}{r_1}\right)\left(1-\frac{2m_0}{r_1}\right)
\left(\frac{g}{r}+{a_\theta}\left(1+\frac{m_0}{r_1}\right)_\theta
-{a}\left(1+\frac{m_0}{r_1}\right)\right.\right.\\\nonumber
&\left.\left.-{2b}\left(1-\frac{m_0}{r_1}\right)\right)\right\}
+\frac{\mu_0}{2\sqrt{2}r}\left(1-\frac{2m_0}{r_1}\right)\left(\frac{2g}{r}+{2a}
\left(1+\frac{m_0}{r_1}\right)-\frac{\bar{Z}}{r^2}\right)\\\nonumber
&\left.+\frac{\mu_0}{2\sqrt{2}}\left(1-\frac{2m_0}{r_1}\right)\left(\frac{g}{r}+{b}
\left(1-\frac{m_0}{r_1}\right)\right)'\right.
\left.-\frac{\chi}{2\sqrt{2}}\left(1-\frac{2m_0}{r_1}\right)
\right.\\\nonumber
&\left.\left\{\frac{3}{2r}+\left(1+\frac{m_0}{r_1}\right)'
\left(1-\frac{m_0}{r_1}\right)\right\}\right..
\end{align}
The quantities introduced in the unstable range in pN limit are
given as
\begin{align*}
\xi_1&={P_0}+\frac{2}{9}
(2{\Pi}_{I0}+{\Pi}_{II0}),\quad\psi=\left(\frac{P_0}{\mu_0+P_0}
+\frac{4\Pi_{I0}}{9(\mu_0+\Pi_{I0})}+
\frac{2\Pi_{II0}}{9(\mu_0+\Pi_{II0})}\right),\\\nonumber
\psi_1&=\frac{7}{4r}+\frac{1}{2}\sigma\left(1-\frac{m_0^2}{r_1^2}
+\frac{1}{r}\right),\quad\psi_2=\left(1-\frac{m_0}{r_1}\right)
\left(1+\frac{3m_0}{r_1}\right),\\\nonumber
\phi_1&=\left(1-\frac{4m_0}{r_1}\right)\left(1
+\frac{m_0}{r_1}\right)\left(1-\frac{3m_0}{r_1}\right)\left(1
-\frac{m_0}{r_1}\right)',\quad\sigma=1-\frac{4m_0^2}{r_1^2},\\\nonumber
\eta_1&={2a}\left( 1+\frac{m_0}{r_1}\right)+{2b}\left(1
-\frac{m_0}{r_1}\right), \quad \phi_4=\left(1
-\frac{m_0}{r_1}\right)_\theta\left(1+\frac{m_0}{r_1}\right)\\\nonumber&
-\frac{\bar{Z}}{r^2},\quad\eta_2=\left(1-\frac{m_0}{r_1}\right)\left(1
+\frac{m_0}{r_1}\right),\quad\phi=\left(1-\frac{m_0}{r_1}\right)\left(1
+\frac{m_0}{r_1}\right)',\\\nonumber
\phi_2&=\left(1-\frac{m_0}{r_1}\right)'\left(1
+\frac{m_0}{r_1}\right),\quad\phi_3=\left(1-\frac{m_0}{r_1}\right)\left(1
+\frac{m_0}{r_1}\right)_\theta,\\\nonumber
\zeta&=\left(a+\frac{am_0}{r_1}\right)'-\frac{\omega\phi_4}{r}
\left(\frac{g}{r}+{a_\theta}\left(1+\frac{m_0}{r_1}\right)_\theta
+\frac{3\bar{Z}}{2r^2}-\eta\right),\\\nonumber
X_1&=-\frac{\mu_0}{4}\left(1-\frac{4m_0}{r_1}\right)\left\{\left(
{a}\left(1+\frac{m_0}{r_1}\right)\right)'-\frac{1}{r}\left(1
-\frac{m_0}{r_1}\right)_\theta\left(1+\frac{m_0}{r_1}\right)
\right.\\\nonumber &\left.\times\left(1-\frac{2m_0}{r_1}\right)
\left({a_\theta}\left(1+\frac{m_0}{r_1}\right)_\theta
-{a}\left(1+\frac{m_0}{r_1}\right)-{2b}\left(1-\frac{m_0}{r_1}\right)\right)\right\}\\\nonumber
&+\frac{\mu_0}{2\sqrt{2}r}\left(1-\frac{2m_0}{r_1}\right)\left\{{2a}
\left(1+\frac{m_0}{r_1}\right)+\left\{{b}
\left(1-\frac{m_0}{r_1}\right)\right\}'\right\}
\left.-\frac{\chi}{2\sqrt{2}}\left(1\right. \right.\\\nonumber
&\left.\left.-\frac{2m_0}{r_1}\right)\left\{\frac{3}{2r}+\left(1+\frac{m_0}{r_1}\right)'
\left(1-\frac{m_0}{r_1}\right)\right\}\right],\\\nonumber
\mathcal{A}_2&=\xi_1\omega_1\left[
\frac{3}{4r}\left(g'+\frac{g}{r}-\frac{\bar{Z}}{2r^2}\right)+\frac{\sigma\eta}{2r^2}
\left(\phi_2+\frac{1}{r}\right)+\frac{1}{4}\left(\eta_1-\frac{\bar{Z}}{r^2}
\right)\right]-2b\omega_1\xi_1\psi_1\\\nonumber
&+\frac{\Pi_{KL0}}{2\sqrt{2}r}\left[6\phi_3
\left\{{b_\theta}\left(1-\frac{m_0}{r_1}\right)_\theta+{b}\left(1-\frac{m_0}{r_1}
\right)\right\}\left({a} \left(1+\frac{m_0}{r_1}\right)
+\frac{c}{r}\right)_\theta\right.\\\nonumber &\left.+2\sigma\eta_1
\left(\eta_1-\frac{\bar{Z}}{r^2}\right)_\theta\right]+\frac{b\chi\phi_1\xi_1}{2}
-{\xi_1\omega_1}\left(\frac{c}{r}\right)'+\frac{1}{2\sqrt{2}}\psi_2\eta\Pi_{KL0,\theta},\\\nonumber
\mathcal{G}_3&=-\frac{\mu_0}{4}\left(1-\frac{4m_0}{r_1}\right)\left\{-\frac{1}{r}\left(1
-\frac{m_0}{r_1}\right)_\theta\left(1+\frac{m_0}{r_1}\right)
\left(1-\frac{2m_0}{r_1}\right)
\left(\frac{g}{r}\right)\right\}+\frac{\mu_0}{2}\\\nonumber
&\times\frac{1}{\sqrt{2}r}\left(1-\frac{2m_0}{r_1}\right)\left(\frac{2g}{r}-\frac{\bar{Z}}{r^2}\right)
+\frac{\mu_0}{2\sqrt{2}}\left(1-\frac{2m_0}{r_1}\right)\left(\frac{g}{r}\right)',\\\nonumber
\mathcal{A}_3&=\omega_1(\psi\chi)'-\psi\psi_1\omega_1+\psi_2
\left(\frac{\Pi_{KL0}\chi}{\mu_0+\Pi_{KL0}}\right)_\theta.
\end{align*}

\end{document}